%% file: Main.tex
\documentclass[10pt,a4paper,twocolumn]{IEEEtran}
\input{packageList.tex}
\input{commandList.tex}

\title{Covariance Matrix Estimation for Massive MIMO}
\begin{document}
	\author{Karthik~Upadhya,~\IEEEmembership{Student Member,~IEEE,}
	Sergiy~A.~Vorobyov,~\IEEEmembership{Fellow,~IEEE}
	\thanks{K. Upadhya and S. A. Vorobyov are with the Department
		of Signal Processing and Acoustics, Aalto University, Espoo - 02150, Finland. (E-mails: karthik.upadhya@aalto.fi, svor@ieee.org).}
	\thanks{This work was supported in part by the Academy of Finland research grant No. 299243.}
	\thanks{\textit{Corresponding Author: Sergiy A. Vorobyov.}}}
	\maketitle	
	
	\input{texFiles/abstract.tex}
	\input{texFiles/introduction.tex}
	\input{texFiles/systemModel.tex}
	\input{texFiles/existingMethods.tex}
	\input{texFiles/proposedMethod.tex}
	\begin{figure*}	
		\begin{minipage}[t]{0.48\textwidth}
			\centering
			\resizebox{0.7\columnwidth}{!}{\input{figs/fig2.tex}}
			\caption{Normalized MSE of channel estimate of an average user vs. $ \numCoherenceBlock $.}
			\label{fig:mseVsN}
		\end{minipage}
		\hspace{\stretch{3}}
		\begin{minipage}[t]{0.48\textwidth}
			\centering
			\resizebox{0.7\columnwidth}{!}{\input{figs/fig1.tex}}
			\caption{Achievable rate in the UL vs. $ \numCoherenceBlock $.}
			\label{fig:rateVsN}
		\end{minipage}	
		\vspace{-0.25cm}
	\end{figure*}
	\begin{figure*}	
		\begin{minipage}[t]{0.48\textwidth}
			\centering
			\resizebox{0.7\columnwidth}{!}{\input{figs/fig3.tex}}
			\caption{MSE of the estimated covariance matrix of an average user vs. $ \numCoherenceBlock $.}
			\label{fig:covMatMseVsN}
		\end{minipage}
		\hspace{\stretch{3}}
		\begin{minipage}[t]{0.48\textwidth}
			\centering
			\resizebox{0.7\columnwidth}{!}{\input{figs/fig4.tex}}
			\caption{Achievable rate in the UL vs. $ \numCoherenceBlock $.}
			\label{fig:rateVsNUniform}
		\end{minipage}	
		\vspace{-0.5cm}
	\end{figure*}
	\input{texFiles/pilotType.tex}
	\input{texFiles/simulationResults.tex}
	\input{texFiles/conclusion}
	\input{texFiles/appendix}

	\bibliographystyle{IEEEtran}
	\bibliography{IEEEabrv,paperBibliography.bib}
\end{document}

%% file: packageList.tex
\usepackage{tikz}
\usepackage{amsmath,amssymb}
\usepackage{pgfplots}
\pgfplotsset{compat=1.7}
\usepackage{amsthm}
\usepackage{cite}
\usepackage[hidelinks]{hyperref}
\usepackage{tabularx}
\usepackage{array,booktabs}
\usepackage{standalone}
\usepackage{cleveref}
\usepackage{enumitem}
\usepackage{pstricks,graphicx}
\usepackage{algpseudocode,algorithmicx,algorithm}
\usepackage{xr}
\usepackage{mdframed}
\usepackage{accsupp}

%% file: commandList.tex
\newcommand{\mbf}[1]{\mathbf{#1}}
\newcommand{\lrf}[1]{\left\{#1\right\}}
\newcommand{\lrc}[1]{\left(#1\right)}
\newcommand{\lrs}[1]{\left[#1\right]}

\newcommand{\indicator}[1]{\mbf{1}_{\lrf{#1}}}
\newcommand{\rth}{\text{th}}

\newcommand{\eye}{\mbf{I}}

\makeatletter
\g@addto@macro\th@remark{\thm@headpunct{\textnormal{:}}}
\makeatother
\theoremstyle{remark}

\newcommand{\Ugtrless}{%
  \mathrel{\kern0pt\mathop{\gtrless}\limits^{\mathcal{U}_\vIdxOne}_{\overline{\mathcal{U}}_\vIdxOne}}%
}

\newcommand{\gtr}{%
  \mathrel{\kern0pt\mathop{<}\limits^{\mathcal{U}_\vIdxOne}}%
}

\newcommand{\complexJ}{\mathrm{j}}
\newcommand{\coherenceTime}{C}
\newcommand{\ulDuration}{C_u}
\newcommand{\dlDuration}{C_d}

\newcommand*{\vIdxOne}{j} 
\newcommand*{\vIdxTwo}{m}
\newcommand*{\sIdxOne}{\ell}
\newcommand*{\sIdxTwo}{k}
\newcommand*{\sIdxThree}{r}
\newcommand*{\sIdxFour}{s}

\newcommand*{\cellIdx}{c}
\newcommand*{\userIdx}{u}

\newcommand{\expectation}{\mathbb{E}}

\newcommand{\ulTxSymbol}{\mbf{s}}
\newcommand{\ulTxSymbolAsync}{\mbf{z}}
\newcommand{\ulTxData}{\mbf{x}}

\newcommand{\channelEstimateLs}{\widehat{\mbf{h}}}

\newcommand{\channelEstimateLmmse}{\widehat{\mbf{h}}^{\mathrm{LMMSE}}}
\newcommand{\covarianceMatrix}{\mbf{R}}
\newcommand{\sumCovarianceMatrix}{\mbf{Q}}

\newcommand{\coherenceBlockIdx}{n}
\newcommand{\nQ}{N_Q}
\newcommand{\nR}{N_R}
\newcommand{\numCellSubset}{T}
\newcommand{\cellSubsetIdx}{t}
\newcommand{\cellSubset}{\mathcal{T}}
\newcommand{\cellSubsetComplement}[1]{\mathcal{T}_{\backslash{#1}}}

\newcommand{\rhoD}[2]
{
	\ifstrempty{#1}
		{
			\ifstrempty{#2}
				{
					\rho
				}
				{
					\rho_{#2}
				}
		}
		{
			\ifstrempty{#2}
				{
					\rho_{#1}
				}
				{
					\rho_{#1,#2}
				}				
		}
}

\newcommand{\rhoP}[2]
{
	\ifstrempty{#1}
		{
			\ifstrempty{#2}
				{
					\lambda
				}
				{
					\lambda_{#2}
				}
		}
		{
			\ifstrempty{#2}
				{
				    \lambda_{#1}
				}
				{	
					\lambda_{#1,#2} 
				}
		}
}

\newcommand{\ulTotPower}[2]
{
	\ifstrempty{#1}
		{
			\ifstrempty{#2}
				{
					\mu
				}
				{
					\mu_{#2}
				}
		}
		{
			\mu_{#1,#2}
		}
}

\newcommand{\spPilot}{\mbf{p}}
\newcommand{\spDualPilot}{\overline{\spPilot}}

\newcommand{\randPhaseRandomVariable}{\Theta}
\newcommand{\randPhaseRealization}{\theta}
\newcommand{\cellSubsetMapping}{\mathcal{M}}
\newcommand{\pilotSubsequenceIdx}{p}
\newcommand{\crossCorrelationChannelEstimate}[2]{\mbf{R}_{\vIdxOne,\widehat{\mbf{h}}_{{#1}{#2}}^{(1)}\widehat{\mbf{h}}_{{#1}{#2}}^{(2)}}}

\newcommand{\tempVariableSp}[1]{\boldsymbol{\alpha}_{\cellIdx\userIdx}^{({#1})}}	
\newcommand{\tempVariableZp}[1]{\boldsymbol{\epsilon}_{\cellIdx\userIdx}^{({#1})}}
\newcommand{\numCoherenceBlock}{N}
\newcommand{\numStationaryBlock}{\tau_{\mathrm{s}}}

\newcommand{\achRate}{\mathrm{SE}}
\newcommand{\prelogFactor}{\eta}
\newcommand{\rxCombiner}{\mbf{v}}

\newcommand{\dataIdx}{t}

\newcommand{\errorCovarianceMatrix}{\mbf{E}}

\newcommand{\logCapacity}{\Psi}
\newcommand{\numExtraSymbol}{\delta}
\newcommand{\dataPilotPower}{\theta}

%% file: texFiles/abstract.tex
\begin{abstract}
	We propose a novel pilot structure for covariance matrix estimation in massive multiple-input multiple-output (MIMO) systems in which each user transmits two pilot sequences, with the second pilot sequence multiplied by a random phase-shift. The covariance matrix of a particular user is obtained by computing the sample cross-correlation of the channel estimates obtained from the two pilot sequences. This approach relaxes the requirement that all the users transmit their uplink pilots over the same set of symbols. We derive expressions for the achievable rate and the mean-squared error of the covariance matrix estimate when the proposed method is used with staggered pilots. The performance of the proposed method is compared with existing methods through simulations.
\end{abstract}


%% file: texFiles/introduction.tex
\section{Introduction}
Massive multiple-input multiple-output (MIMO) is a variation of multi-user MIMO (MU-MIMO) that has a large  number of antennas at the base station (BS), which significantly improves the spectral efficiency through spatial multiplexing \cite{Marzetta2010Noncooperative, larsson2014massive, lulu2014anoverview, rusek2013scaling} at a low cost of simple linear processing at the BS \cite{Marzetta2010Noncooperative, yang2013performance, hoydis2013massive}. However, in practice, the BS needs to obtain channel state information (CSI) using pilots which have to be reused in different cells, thereby causing pilot contamination \cite{Marzetta2010Noncooperative}. It has been argued that pilot contamination, in independent and identically distributed (i.i.d) Rayleigh fading, puts a fundamental limit on the asymptotically achievable rate in massive MIMO systems \cite{Marzetta2010Noncooperative}, and pilot decontamination algorithms have been designed in many works (see \cite{Muller2014Blind,bjornson2015massive,Yin2013Coordinated,upadhya2015superimposed,upadhya2016superimposed,jose2011pilot} to mention just a few). 

It has been recently shown in \cite{bjornson2017unlimited} that the ceiling on the UL and DL rates due to pilot contamination can be eliminated under certain loose conditions on the covariance matrices of the users. However, this method requires estimates of the covariance matrix at the BS, which have to be obtained from observations that are made in the presence of pilot contamination. In \cite{bjornson2016imperfect}, two methods have been developed wherein the users are assigned unique pilots specifically for estimating the covariance matrices. In \cite{neumann2017jointCovariance}, a method for jointly performing pilot allocation and estimating the covariance matrix has been proposed. A method for estimating a low-rank covariance matrix has also been proposed in \cite{caire2017massive}. A common theme in all the earlier works is that they require/assume that the users in all cells transmit their UL pilots simultaneously, which can be infeasible in practice.

In this letter, we develop a method for estimating the users' covariance matrices using a pair of pilot sequences, with the second pilot sequence multiplied by a random phase-shift. The quality of the covariance estimate obtained through the proposed method using staggered pilots is quantified using expressions for its mean-squared error (MSE). The achievable rate is analyzed numerically when the covariance matrices are estimated and used with staggered and regular pilots. 

\textit{\textbf{Notation}}: A vector is denoted as $ \mbf{a} $ and a matrix as $ \mbf{A} $. The notations $ \lrc{\cdot}^T $, $ \lrc{\cdot}^* $, $ \lrc{\cdot}^H $, $ \lrc{\cdot}^{-1} $, $ \mathrm{blkdiag}\lrf{\cdot} $, $ \mathrm{trace}\lrc{\cdot} $, $ \lfloor x \rfloor $ represent the transpose, conjugate, conjugate transpose, matrix inverse, block diagonal matrix, matrix trace, and largest integer smaller than $ x $, respectively, while $ \mathcal{CN}(\boldsymbol{\mu},\mbf{\Sigma}) $ stands for the complex Gaussian distribution with mean $ \boldsymbol{\mu} $ and covariance matrix $ \mbf{\Sigma} $.


%% file: texFiles/systemModel.tex
\section{System Model and Problem Description}
\label{sec:systemModel}
We consider a massive MU-MIMO system with $ L $ cells each having $ M $ antennas at the BS and containing $ K $ users. Denoting a user $ \userIdx $ in cell $ \cellIdx $ as $ \lrc{\cellIdx,\userIdx} $, the channel vector between user $ \lrc{\cellIdx,\userIdx} $ and BS $ \vIdxOne $ is represented as $ \mbf{h}_{\vIdxOne\cellIdx\userIdx} \in \mathbb{C}^{M} $ and is distributed as $ \mathcal{CN}\lrc{\mbf{0},\covarianceMatrix_{\vIdxOne\cellIdx\userIdx}} $.
The channel is assumed to be constant for $ \coherenceTime $ symbols and the second-order statistics $ \mbf{R}_{\vIdxOne\cellIdx\userIdx} $ are assumed to be constant for $ \tau_s $ blocks each containing $ \coherenceTime $ symbols. The coherence time of the channel is divided into $ \ulDuration $ and $ \dlDuration $ symbols for the UL and DL time-slots, respectively.

In \cite{bjornson2017unlimited} and \cite{bjornson2017pilot}, it was shown that the UL and DL rates increase asymptotically in $ M $ when the data is estimated using a linear minimum mean-squared error (LMMSE) or zero-forcing (ZF) precoder/combiner that is designed using the LMMSE channel estimate.
If $ \channelEstimateLs_{\vIdxOne\cellIdx\userIdx} $ is the least-square (LS) estimate of the channel, the channel estimate obtained using the LMMSE criterion can be written as
\begin{align}
	\channelEstimateLmmse_{\vIdxOne\cellIdx\userIdx}
	&=
	\expectation
	\lrf{
	\mbf{h}_{\vIdxOne\cellIdx\userIdx}
	\mbf{h}_{\vIdxOne\cellIdx\userIdx}^H
	}
		\expectation\lrf{
				\channelEstimateLs_{\vIdxOne\vIdxOne\userIdx}
				\channelEstimateLs_{\vIdxOne\vIdxOne\userIdx}^H
		}^{-1}
	\channelEstimateLs_{\vIdxOne\vIdxOne\userIdx}
	\nonumber\\
	&=
	\covarianceMatrix_{\vIdxOne\cellIdx\userIdx}
	\sumCovarianceMatrix_{\vIdxOne\userIdx}^{-1}
	\channelEstimateLs_{\vIdxOne\vIdxOne\userIdx}
	\label{eqn:lmmseChannelEstimate}
\end{align}
where $ \sumCovarianceMatrix_{\vIdxOne\userIdx	} \triangleq \expectation\lrf{\channelEstimateLs_{\vIdxOne\vIdxOne\userIdx}\channelEstimateLs_{\vIdxOne\vIdxOne\userIdx}^H}$ and $ \covarianceMatrix_{\vIdxOne\cellIdx\userIdx} \triangleq \expectation\lrf{\mbf{h}_{\vIdxOne\cellIdx\userIdx}\mbf{h}_{\vIdxOne\cellIdx\userIdx}^H} $. 

Utilizing \eqref{eqn:lmmseChannelEstimate}, the corresponding multi-cell LMMSE combining vector is
\begin{equation}
	\mbf{v}_{\vIdxOne\userIdx}
	=
	\lrc{
		\sum\limits_{\sIdxOne=0}^{L-1}
		\sum\limits_{\sIdxTwo=0}^{K-1}
		\channelEstimateLmmse_{\vIdxOne\sIdxOne\sIdxTwo}
		\lrc{\channelEstimateLmmse_{\vIdxOne\sIdxOne\sIdxTwo}}^H
		\!\!\!\!\!
		+
		\!
		\mbf{Z}_{\vIdxOne}}^{-1}\!\!\!\!\!\!
	\channelEstimateLmmse_{\vIdxOne\vIdxOne\userIdx}
	\label{eqn:mmseCombiner}
\end{equation}
where $ \mbf{Z}_{\vIdxOne} \triangleq \sum\limits_{\sIdxOne=0}^{L-1}\sum\limits_{\sIdxTwo=0}^{K-1} \lrc{\covarianceMatrix_{\vIdxOne\sIdxOne\sIdxTwo} - \covarianceMatrix_{\vIdxOne\sIdxOne\sIdxTwo} \sumCovarianceMatrix_{\vIdxOne\sIdxTwo}^{-1} \covarianceMatrix_{\vIdxOne\sIdxOne\sIdxTwo} } + \sigma^2 \eye_M$. From \eqref{eqn:lmmseChannelEstimate} and \eqref{eqn:mmseCombiner}, it can be observed that obtaining the LMMSE estimate of the channel and data requires the covariance matrices $ \covarianceMatrix_{\vIdxOne\sIdxOne\sIdxTwo},\; \forall\; \sIdxOne,\sIdxTwo $ and $ \sumCovarianceMatrix_{\vIdxOne\sIdxTwo},\; \forall\; \sIdxTwo $. In practice, $ \covarianceMatrix_{\vIdxOne\sIdxOne\sIdxTwo} $ has to be estimated in the presence of pilot contamination, which complicates the estimation problem since the estimate $ \widehat{\covarianceMatrix}_{\vIdxOne\sIdxOne\sIdxTwo} $ is contaminated by the covariance matrices of users in adjacent cells that employ the same pilot.

%% file: texFiles/existingMethods.tex

Existing methods for covariance matrix estimation employ regular pilots for both channel and covariance matrix estimation wherein the channel and covariance matrix estimates are obtained by dedicating a part of the time-frequency resource for pilot transmission. Under the assumption that the pilot transmission from all the cells are synchronized\footnote{Pilots which are transmitted simultaneously by the users in all cells are henceforth referred to as regular pilots.} and that every cell transmits the same pilots, the received observations at BS $ \vIdxOne $ during pilot transmission in the $ \coherenceBlockIdx\rth $ coherence block can be written as
\begin{align}
	\mbf{Y}_{\vIdxOne}^{\lrc{\coherenceBlockIdx}}
	&=
	\sum\limits_{\sIdxOne=0}^{L-1}
	\sum\limits_{\sIdxTwo=0}^{K-1}
	\sqrt{\ulTotPower{}{}}
	\mbf{h}_{\vIdxOne\sIdxOne\sIdxTwo}^{\lrc{\coherenceBlockIdx}}
	\pmb{\phi}_{\sIdxTwo}^T
	+
	\mbf{W}_{\vIdxOne}^{\lrc{\coherenceBlockIdx}}
\end{align}
where $ \mbf{Y}_{\vIdxOne}^{\lrc{\coherenceBlockIdx}} \in \mathbb{C}^{M\times\tau} $ are the received observations, $ \pmb{\phi}_{\sIdxTwo} \in \mathbb{C}^{\tau} $ is the pilot sequence transmitted by user $ \sIdxTwo $, $ \ulTotPower{}{} $ is the uplink transmit power, and $ \mbf{W}_{\vIdxOne}^{\lrc{\coherenceBlockIdx}} \in \mathbb{C}^{M\times\tau} $ is the additive noise at the BS with each element i.i.d as $ \mathcal{CN}(0,\sigma^2) $. Assuming that the pilots $ \pmb{\phi}_{\sIdxTwo} $ are taken from the columns of a scaled unitary matrix $ \mbf{\Phi} \in \mathbb{C}^{\tau\times\tau} $ with $ \mbf{\Phi}^H \mbf{\Phi} = \tau \eye_{\tau} $, the LS estimate of the channel can be obtained as
\begin{align}
	\channelEstimateLs_{\vIdxOne\cellIdx\userIdx}^{\lrc{\coherenceBlockIdx}}\!\!
	&=
	\frac{1}{\tau\sqrt{\ulTotPower{}{}}}
	\mbf{Y}_{\vIdxOne}^{\lrc{\coherenceBlockIdx}}
	\pmb{\phi}_{\userIdx}^*
	=
	\mbf{h}_{\vIdxOne\cellIdx\userIdx}^{\lrc{\coherenceBlockIdx}}
	+
	\sum\limits_{\sIdxOne\neq\vIdxOne}
	\mbf{h}_{\vIdxOne\sIdxOne\userIdx}^{\lrc{\coherenceBlockIdx}}
	+
	\frac{\mbf{W}^{\lrc{\coherenceBlockIdx}}_{\vIdxOne}\pmb{\phi}_{\vIdxTwo}^*}{\tau\sqrt{\ulTotPower{}{}}}	\;.
\end{align}
Since $ \mbf{Q}_{\vIdxOne\userIdx} = \expectation\lrf{\widehat{\mbf{h}}_{\vIdxOne\vIdxOne\userIdx}\widehat{\mbf{h}}_{\vIdxOne\vIdxOne\userIdx}^H} $, its estimate
can be obtained from the sample mean of $ \channelEstimateLs_{\vIdxOne\vIdxOne\userIdx}^{\lrc{\coherenceBlockIdx}} $ over $ N_Q $ coherence blocks as
\begin{align}
	\widehat{\sumCovarianceMatrix}_{\vIdxOne\userIdx}
	=
	\frac{1}{\nQ}
	\sum\limits_{\coherenceBlockIdx=0}^{\nQ-1}
	\channelEstimateLs_{\vIdxOne\vIdxOne\userIdx}^{\lrc{\coherenceBlockIdx}}
	\lrc{\channelEstimateLs_{\vIdxOne\vIdxOne\userIdx}^{\lrc{\coherenceBlockIdx}}}^H \;.
\end{align}
		However, as mentioned earlier, estimating individual covariance matrices $ \mbf{R}_{\vIdxOne\sIdxOne\sIdxTwo},  \; \forall \sIdxOne,\sIdxTwo  $ is challenging, since channel observations are made in the presence of pilot contamination. In \cite{bjornson2016imperfect}, $ \mbf{R}_{\vIdxOne\sIdxOne\sIdxTwo} $ is estimated indirectly through $ {\sumCovarianceMatrix_{\vIdxOne\sIdxOne,-\sIdxTwo} \triangleq \sumCovarianceMatrix_{\vIdxOne\sIdxTwo}-\covarianceMatrix_{\vIdxOne\sIdxOne\sIdxTwo}} $ which is the sum covariance matrix of the channels of all the interfering users using the same pilot as user $ \lrc{\sIdxOne,\sIdxTwo} $. $ \widehat{\sumCovarianceMatrix}_{\vIdxOne\sIdxOne,-\sIdxTwo} $ is estimated separately using $ \nR $ unique orthogonal pilots for each $ \sIdxTwo $ and then subtracted from $ \widehat{\sumCovarianceMatrix}_{\vIdxOne\sIdxTwo} $ to obtain $ \widehat{\covarianceMatrix}_{\vIdxOne\sIdxOne\sIdxTwo} $, i.e.,

\begin{equation}
	\widehat{\covarianceMatrix}_{\vIdxOne\sIdxOne\sIdxTwo}
	=
	\widehat{\sumCovarianceMatrix}_{\vIdxOne\sIdxTwo}
	-
	\widehat{\sumCovarianceMatrix}_{\vIdxOne\sIdxOne,-\sIdxTwo} \;.
\end{equation}
When $ M $ is larger than $ \nQ $ and $ \nR $, the resulting estimates of $ \widehat{\covarianceMatrix}_{\vIdxOne\sIdxOne\sIdxTwo} $ and $ \widehat{\sumCovarianceMatrix}_{\vIdxOne\sIdxTwo} $ have to be regularized in order to ensure full-rank and positive (semi-)definiteness \cite{bjornson2016imperfect}.
For a massive MIMO system with $ L $ cells and $ K $ users per cell, estimating both $ \covarianceMatrix_{\vIdxOne\sIdxOne\sIdxTwo}, \; \forall \sIdxOne,\sIdxTwo $ and $ \sumCovarianceMatrix_{\vIdxOne\sIdxTwo} $ using this approach would require $ LK\nR + K\nQ $ UL training symbols. In addition, utilizing unique pilots for estimating $ \widehat{\sumCovarianceMatrix}_{\vIdxOne\sIdxOne,-\sIdxTwo} $ implicitly assumes and requires that the users in all the $ L $ cells transmit UL pilots simultaneously. While such an assumption is common in massive MIMO literature, it may not be practically feasible since it requires that the BSs coordinate the UL pilot transmissions of their users.

%% file: texFiles/proposedMethod.tex
\section{Proposed Pilot Structure and Method for Estimating Covariance Matrices}
\label{sec:proposedMethod}
In the proposed approach, we assume that the $ L $ cells are divided into $ \numCellSubset $ subsets with the $ \cellSubsetIdx\rth $ subset containing $ L_{\cellSubsetIdx} $ contiguous cells. Here $ L_{\cellSubsetIdx} $ is chosen such that $ 1\leq L_{\cellSubsetIdx}\leq\lfloor\ulDuration/2K\rfloor $. The $ L_{\cellSubsetIdx} $ cells within each of the $ \numCellSubset $ subsets are assumed to be able to coordinate their UL pilot transmissions, whereas the cells in two different subsets transmit pilot and data asynchronously. Let $ {\cellSubsetMapping_{} : \lrf{1,\ldots,L}\rightarrow\lrf{1,\ldots,T}} $ be the mapping between a cell and its corresponding subset, and let $ \cellSubsetIdx = \cellSubsetMapping_{\sIdxOne} $.

User $ \lrc{\sIdxOne,\sIdxTwo} $ transmits the symbol vector $ \ulTxSymbol_{\sIdxOne\sIdxTwo}^{(\coherenceBlockIdx)} \triangleq \rhoD{}{} \ulTxData_{\sIdxOne\sIdxTwo}^{(\coherenceBlockIdx)} + \rhoP{}{} \spDualPilot_{\sIdxOne\sIdxTwo}^{(\coherenceBlockIdx)}  $ in the UL in the $ \coherenceBlockIdx\rth $ coherence block, where $ \spDualPilot_{\sIdxOne\sIdxTwo}^{(\coherenceBlockIdx)} \in \mathbb{C}^{\ulDuration} $ is the UL pilot, $ \ulTxData_{\sIdxOne\sIdxTwo}^{(\coherenceBlockIdx)} \in \mathbb{C}^{\ulDuration} $ is the UL data, and $ \rhoD{}{}^2 $ and  $ \rhoP{}{}^2 $ are the fractions of power with which data and pilots are transmitted, respectively. Then, we assume that the pilot sequence $ \spDualPilot_{\sIdxOne\sIdxTwo}^{(\coherenceBlockIdx)} $ is comprised of two subsequences and can be written as
\begin{align}
	\spDualPilot_{\sIdxOne\sIdxTwo}^{(\coherenceBlockIdx)}
	\triangleq
	\lrs{\spPilot_{\sIdxOne\sIdxTwo}^T,e^{\complexJ\randPhaseRealization_{\cellSubsetIdx,\coherenceBlockIdx}}\spPilot_{\sIdxOne\sIdxTwo}^T}^T
\end{align} 
where $ \lrf{\randPhaseRealization_{\cellSubsetIdx,\coherenceBlockIdx}}_{\coherenceBlockIdx=1}^{N} $ are $ N $ realizations of a random variable $ \randPhaseRandomVariable_{\cellSubsetIdx} $. The random variable $ \randPhaseRandomVariable_{\cellSubsetIdx} $ is assumed to be independent of the channel and data vectors and distributed such that $ \lrf{\randPhaseRandomVariable_{\cellSubsetIdx}}_{\cellSubsetIdx=1}^{\numCellSubset} $ are mutually independent and $ \expectation\lrs{e^{\complexJ\randPhaseRandomVariable_{\cellSubsetIdx}}}=0, \; \forall\cellSubsetIdx $. In addition, we also assume that $ \randPhaseRealization_{\cellSubsetIdx,\coherenceBlockIdx}, \;\forall\cellSubsetIdx,\coherenceBlockIdx$ are known to all the $ L $ BSs, and that the subsequence $ \spPilot_{\sIdxOne\sIdxTwo} $ is chosen from the columns of a scaled unitary matrix $ \mbf{P} $, where $ \mbf{P} $ is such that $ \mbf{P}^H\mbf{P} = K\eye_{\ulDuration/2} $. Also, since $ \ulDuration \geq 2 L_{\cellSubsetIdx} K $, each user in the $ L_{\cellSubsetIdx} $ cells from subset $ \cellSubsetIdx $ can be assigned a unique pilot $ \spDualPilot_{\sIdxOne\sIdxTwo} $. It has to be noted that the symbol vector $ \ulTxSymbol_{\sIdxOne\sIdxTwo}^{(\coherenceBlockIdx)} $ can either contain regular\footnote{Note that with regular pilots, users in all the $ L $ cells transmit pilots simultaneously and the condition $ \ulDuration \geq 2 L_{\cellSubsetIdx} K $ does not apply.}, staggered, or superimposed pilots depending on the contents of $ \spPilot_{\sIdxOne\sIdxTwo} $ and $ \ulTxData_{\sIdxOne\sIdxTwo}^{(\coherenceBlockIdx)} $. With staggered pilots, the users in different cells stagger their UL pilot transmissions \cite{mahyiddin2015performance,kong2016Channel}, and with superimposed pilots, the users transmit UL pilots alongside data \cite{upadhya2016superimposed}.

Let $ \mbf{Y}_{\vIdxOne}^{\lrc{\coherenceBlockIdx,\pilotSubsequenceIdx}} \in \mathbb{C}^{M\times\ulDuration/2} $ for $ \pilotSubsequenceIdx = 1,2 $ be the received observations at BS $ \vIdxOne $ when the first and second pilot subsequences are transmitted in the $ \coherenceBlockIdx\rth $ coherence block. Then, $ \mbf{Y}_{\vIdxOne}^{\lrc{\coherenceBlockIdx,\pilotSubsequenceIdx}}, \; \forall p=1,2 $ can be written as
\begin{align}
	\mbf{Y}_{\vIdxOne}^{\lrc{\coherenceBlockIdx,\pilotSubsequenceIdx}}
	=
	\sum\limits_{\sIdxOne=0}^{L-1}
	\sum\limits_{\sIdxTwo=0}^{K-1}
	\sqrt{\ulTotPower{}{}}
	\mbf{h}_{\vIdxOne\sIdxOne\sIdxTwo}
	\lrc{\ulTxSymbol^{\lrc{\coherenceBlockIdx,\pilotSubsequenceIdx}}_{\sIdxOne\sIdxTwo}}^T
	+
	\mbf{W}_{\vIdxOne}^{\lrc{\coherenceBlockIdx,\pilotSubsequenceIdx}}
	\label{eqn:receivedObservations}
	\end{align}
where $ \ulTxSymbol^{\lrc{\coherenceBlockIdx,\pilotSubsequenceIdx}}_{\sIdxOne\sIdxTwo} $ and $ \mbf{W}_{\vIdxOne}^{\lrc{\coherenceBlockIdx,\pilotSubsequenceIdx}} $ are the transmitted symbols and additive noise at the BS during the transmission of the $ \pilotSubsequenceIdx\rth $ pilot subsequence. 

Dropping the index $ \coherenceBlockIdx $, for an arbitrary user $ \userIdx $ in cell $ \cellIdx $ at BS $ \vIdxOne $, consider the cross correlation between the LS estimates of the channel obtained from the first and second pilot subsequences, that is,
\begin{align}
	\crossCorrelationChannelEstimate{\cellIdx}{\userIdx}
	&\triangleq\expectation
	\left[
			\lrf{
					\mbf{Y}_{\vIdxOne}^{\lrc{1}}
					\lrc{
						\rhoP{}{}^2
						\spPilot_{\cellIdx\userIdx}^T
						\spPilot_{\cellIdx\userIdx}^*
					}^{-1}
					\rhoP{}{}						
					\spPilot_{\cellIdx\userIdx}^*
				}
			\right.
			\nonumber\\
			&
			\left.\times
			\lrf{				
				\mbf{Y}_{\vIdxOne}^{\lrc{2}}
				\lrc{
					\rhoP{}{}^2
					\spPilot_{\cellIdx\userIdx}^T
					\spPilot_{\cellIdx\userIdx}^*
				}^{-1}
				\rhoP{}{}				
				e^{-\complexJ\randPhaseRandomVariable_{\cellSubsetMapping_{\cellIdx}}}		
				\spPilot_{\cellIdx\userIdx}^*
				}^H
		\right]\;.
		\label{eqn:channelCrossCorrelationDefn}
\end{align}
Substituting \eqref{eqn:receivedObservations} and the definition of $ \ulTxSymbol_{\cellIdx\userIdx} $ into \eqref{eqn:channelCrossCorrelationDefn}, we obtain
\begin{align}
	&\crossCorrelationChannelEstimate{\cellIdx}{\userIdx}
	=
	\expectation
	\left[
			\left\{
					\mbf{h}_{\vIdxOne\cellIdx\userIdx}
					+										
					\tempVariableSp{1}
					+										
					\tempVariableZp{1}
					+
					\mbf{w}_{\vIdxOne}^{(1)}
				\right\}
				\times
				\Big\{
				\mbf{h}_{\vIdxOne\cellIdx\userIdx}					
				\right.
				\nonumber
				\\
				&\!\!\!\left.
				\!
					+
					e^{-\complexJ\randPhaseRandomVariable_{\cellSubsetMapping_{\cellIdx}}}		
					\tempVariableSp{2}
					\!
					\!
					+
					\!
					e^{-\complexJ\randPhaseRandomVariable_{\cellSubsetMapping_{\cellIdx}}}		
					\tempVariableZp{2}
					\!
					+
					\!
					e^{-\complexJ\randPhaseRandomVariable_{\cellSubsetMapping_{\cellIdx}}}
					\mbf{w}_{\vIdxOne}^{(2)}
			\Big\}^{\!\!H}
		\right]
		\!\!
		=
		\covarianceMatrix_{\vIdxOne\cellIdx\userIdx}		
		\label{eqn:channelCrossCorrelation}
\end{align}
where, for $ \pilotSubsequenceIdx = 1,2 $, we have
\begin{align}
	\tempVariableSp{\pilotSubsequenceIdx}
	&\triangleq
	\frac{\rhoD{}{}}{K\rhoP{}{}}
	\sum\limits_{\sIdxOne\in\cellSubset_{\cellIdx}}
	\sum\limits_{\sIdxTwo=0}^{K-1}
	\mbf{h}_{\vIdxOne\sIdxOne\sIdxTwo}	
	\lrc{\ulTxData_{\sIdxOne\sIdxTwo}^{\lrc{\pilotSubsequenceIdx}}}^T
	\spPilot_{\cellIdx\userIdx}^*	
	\\
	\tempVariableZp{\pilotSubsequenceIdx}
	&\triangleq
	\frac{1}{K\rhoP{}{}}
	\sum\limits_{\sIdxOne\in\cellSubsetComplement{\cellIdx}}
	\sum\limits_{\sIdxTwo=0}^{K-1}
	\mbf{h}_{\vIdxOne\sIdxOne\sIdxTwo}	
	\lrc{\ulTxSymbolAsync_{\sIdxOne\sIdxTwo}^{\lrc{\pilotSubsequenceIdx}}}^T
	\spPilot_{\cellIdx\userIdx}^*
	\\
	\mbf{w}_{\vIdxOne}^{(\pilotSubsequenceIdx)}	
	&\triangleq
	\mbf{W}_{\vIdxOne}^{(\pilotSubsequenceIdx)}\spPilot_{\cellIdx\userIdx}^* / \lrc{K\rhoP{}{}\sqrt{\ulTotPower{}{}}}\;.
\end{align}
Here $ \cellSubset_{\cellIdx} = \lrf{\sIdxOne\;|\;\cellSubsetMapping_{\sIdxOne}=\cellSubsetMapping_{\cellIdx}} $ is the set of cells that are in the same subset as cell $ \cellIdx $ and $ \cellSubsetComplement{\cellIdx} $ is its complement, $ \ulTxSymbolAsync_{\sIdxOne\sIdxTwo} $ is the vector of symbols (either pilots or data) transmitted asynchronously by a user\footnote{If a cell in $ \cellSubsetComplement{\cellIdx} $ is transmitting in the DL, each BS antenna is treated as a user.} in $ \cellSubsetComplement{\cellIdx} $. In \eqref{eqn:channelCrossCorrelation}, $ e^{\complexJ\randPhaseRandomVariable_{\cellSubsetMapping_{\cellIdx}}} $ decorrelates the channel estimation errors resulting from the transmissions from the users in $ \cellSubsetComplement{\cellIdx} $, which in turn causes the cross-correlation of the channel estimates to become equal to $ \covarianceMatrix_{\vIdxOne\cellIdx\userIdx} $.

Using the result in \eqref{eqn:channelCrossCorrelation}, an estimate of $ \covarianceMatrix_{\vIdxOne\cellIdx\userIdx} $ can be obtained by the sample cross-correlation of both the channel estimates averaged over $ \numCoherenceBlock $ coherence blocks, i.e., 
\begin{align}
	\widehat{\covarianceMatrix}_{\vIdxOne\cellIdx\userIdx}
	&=
	\frac{1}{\numCoherenceBlock}
	\sum\limits_{\coherenceBlockIdx=1}^{\numCoherenceBlock}
	\widehat{\mbf{h}}_{\vIdxOne\cellIdx\userIdx}^{\lrc{\coherenceBlockIdx,1}}
	\lrc{\widehat{\mbf{h}}_{\vIdxOne\cellIdx\userIdx}^{\lrc{\coherenceBlockIdx,2}}}^H
\end{align}
where
\begin{align}
	\widehat{\mbf{h}}_{\vIdxOne\cellIdx\userIdx}^{\lrc{\coherenceBlockIdx,1}}
	&=
	\mbf{Y}_{\vIdxOne}^{\lrc{\coherenceBlockIdx,1}}
	\lrc{
		\rhoP{}{}^2
		\spPilot_{\cellIdx\userIdx}^T
		\spPilot_{\cellIdx\userIdx}^*
	}^{-1}
	\rhoP{}{}						
	\spPilot_{\cellIdx\userIdx}^*
	\\
	\widehat{\mbf{h}}_{\vIdxOne\cellIdx\userIdx}^{\lrc{\coherenceBlockIdx,2}}
	&=
	\mbf{Y}_{\vIdxOne}^{\lrc{\coherenceBlockIdx,2}}
	\lrc{
		\rhoP{}{}^2
		\spPilot_{\cellIdx\userIdx}^T
		\spPilot_{\cellIdx\userIdx}^*
	}^{-1}
	\rhoP{}{}
	e^{-\complexJ\randPhaseRealization_{\cellSubsetMapping_{\cellIdx},\coherenceBlockIdx}}
	\spPilot_{\cellIdx\userIdx}^*\;.
\end{align}
It is straightforward to show that the sample cross-correlation converges in probability to the true correlation, i.e., 
$
	\widehat{\covarianceMatrix}_{\vIdxOne\cellIdx\userIdx}
	\xrightarrow[\numCoherenceBlock\rightarrow\infty]{P}
	\covarianceMatrix_{\vIdxOne\cellIdx\userIdx}\;.
$

However, for a finite $ \numCoherenceBlock $, the estimate $ \widehat{\covarianceMatrix}_{\vIdxOne\cellIdx\userIdx} $ is not necessarily Hermitian symmetric. Therefore, this matrix can be regularized by approximating it with a positive semi-definite matrix. Thus, we approximate $ \widehat{\covarianceMatrix}_{\vIdxOne\cellIdx\userIdx} $ with the positive semidefinite matrix closest in Frobenius norm, which can be easily shown to be
$
	\widehat{\covarianceMatrix}_{\vIdxOne\cellIdx\userIdx}^{\mathrm{PSD}}
	\triangleq
	\mbf{U}
	\mbf{D}_{+}
	\mbf{U}^H
$
where $ \mbf{D}_{+} $ is a diagonal matrix that contains only the positive eigenvalues of the symmetric part of $ \widehat{\covarianceMatrix}_{\vIdxOne\cellIdx\userIdx} $, i.e., $ \widehat{\covarianceMatrix}_{\vIdxOne\cellIdx\userIdx}^{\mathrm{sym}}\triangleq \lrc{\widehat{\covarianceMatrix}_{\vIdxOne\cellIdx\userIdx} + \widehat{\covarianceMatrix}_{\vIdxOne\cellIdx\userIdx}^H}/2 $, and $ \mbf{U} $ contains the corresponding eigenvectors.

In order to estimate the channel covariance matrix of a user $ \lrc{\cellIdx,\userIdx} $ at an arbitrary BS $ \vIdxOne $, the BS requires only the knowledge of $ \lrf{\randPhaseRealization_{\cellSubsetMapping_{\cellIdx},\coherenceBlockIdx}}_{\coherenceBlockIdx=1}^{\numCoherenceBlock} $ and the symbol and subcarrier indices in which user $ \lrc{\cellIdx,\userIdx} $ transmits its UL pilots. As a result, unlike in \cite{bjornson2016imperfect} and \cite{neumann2017jointCovariance}, the proposed method does not require that all the users transmit pilots simultaneously\footnote{The proposed method does not require	the users in different cells to transmit UL pilots over the same set of UL symbols. However, note that we are still assuming symbol-level synchronization over all the cells.}. In fact, as will be shown in Section \ref{sec:simulationResults}, the proposed method performs well even when $ L_{\cellSubsetIdx} = 1,\;\forall\cellSubsetIdx $, i.e., when none of the BSs coordinate the UL pilot/data transmissions of their users. 

%% file: figs/fig2.tex
\begin{tikzpicture}[scale = 1]

\pgfplotsset{every axis legend/.append style={at={(0.35,0.7)},anchor=west},legend cell align = {left}}
        
\pgfplotsset{%
tick label style={font=\Large},
title style={font=\normalsize,align=center},
label style={font=\Large,align=center},
legend style={font=\Large}
            }

\begin{axis}[%
width=4.822in,
height=3.703in,
at={(0.809in,0.513in)},
xlabel = {$ \numCoherenceBlock $},
ylabel = {Normalized MSE of the Channel Estimate (dB)},
xmin = 0,
xmax = 2500,
xtick = {0,500,1000,1500,2000,2500},
ymin = -14,
ymax = 0,
cycle list name = {color},
grid = major,
scale only axis,
scale = 1]

\addlegendentry{LS}

\addplot [ color = {cyan}, 
mark  = {diamond},
mark options = {solid},
style = {solid},
line width = 1pt] table {figs/datFiles/fig2-line4.dat};

\addlegendentry{Method in \cite{bjornson2016imperfect}}

\addplot [ color = {black}, 
mark  = {o},
mark options = {solid},
style = {solid},
line width = 1pt] table {figs/datFiles/fig2-line3.dat};

\addlegendentry{Proposed Method - $ L_{\cellSubsetIdx}=1 $}

\addplot [ color = {magenta}, 
mark  = {square},
mark options = {solid},
style = {solid},
line width = 1pt] table {figs/datFiles/fig2-line5.dat};

\addlegendentry{Proposed Method - Staggered Pilot}

\addplot [ color = {blue}, 
           mark  = {asterisk},
           mark options = {solid},
           style = {solid},
           line width = 1pt] table {figs/datFiles/fig2-line1.dat};

\addlegendentry{LMMSE}

\addplot [ color = {red}, 
           mark  = {none},
           mark options = {solid},
           style = {solid},
           line width = 1pt] table {figs/datFiles/fig2-line2.dat};

\end{axis}
\end{tikzpicture}

%% file: figs/fig1.tex
\begin{tikzpicture}[scale = 1]

\pgfplotsset{every axis legend/.append style={at={(0.35,0.4)},anchor=west},legend cell align = {left}}
        
\pgfplotsset{%
tick label style={font=\Large},
title style={font=\normalsize,align=center},
label style={font=\Large,align=center},
legend style={font=\Large}
            }

\begin{axis}[%
width=4.822in,
height=3.703in,
at={(0.809in,0.513in)},
xlabel = {$ \numCoherenceBlock $},
ylabel = {Avg. Sum Rate},
xtick = {0,500,1000,1500,2000,2500},
xmin = 0,
xmax = 2500,
ymin = 12,
ymax = 30,
cycle list name = {color},
grid = major,
scale only axis,
scale = 1]

\addlegendentry{LMMSE}

\addplot [ color = {red}, 
mark  = {none},
mark options = {solid},
style = {solid},
line width = 1pt] table {figs/datFiles/fig1-line2.dat};

\addlegendentry{Proposed Method - Staggered Pilot}

\addplot [ color = {blue}, 
           mark  = {asterisk},
           mark options = {solid},
           style = {solid},
           line width = 1pt] table {figs/datFiles/fig1-line1.dat};

\addlegendentry{Proposed Method - $ L_{\cellSubsetIdx} = 1 $}

\addplot [ color = {magenta}, 
mark  = {square},
mark options = {solid},
style = {solid},
line width = 1pt] table {figs/datFiles/fig1-line5.dat};

\addlegendentry{Method in \cite{bjornson2016imperfect}}

\addplot [ color = {black}, 
           mark  = {o},
           mark options = {solid},
           style = {solid},
           line width = 1pt] table {figs/datFiles/fig1-line3.dat};

\addlegendentry{LS}

\addplot [ color = {cyan}, 
           mark  = {diamond},
           mark options = {solid},
           style = {solid},
           line width = 1pt] table {figs/datFiles/fig1-line4.dat};

\end{axis}
\end{tikzpicture}

%% file: figs/fig3.tex
\begin{tikzpicture}[scale = 1]

\pgfplotsset{every axis legend/.append style={at={(0.3,0.55)},anchor=west},legend cell align = {left}}
        
\pgfplotsset{%
tick label style={font=\Large},
title style={font=\normalsize,align=center},
label style={font=\Large,align=center},
legend style={font=\Large}
            }

\begin{axis}[%
width=4.822in,
height=3.703in,
at={(0.809in,0.513in)},
xlabel = {$ \numCoherenceBlock $},
ylabel = {MSE of the Covariance Matrix Estimate (dB)},
cycle list name = {color},
grid = major,
xtick = {0,500,1000,1500,2000,2500},
scale only axis,
xmin = 0,
xmax = 2500,
ymin = 10,
ymax = 35,
scale = 1]

\addlegendentry{Method in \cite{bjornson2016imperfect}}

\addplot [ color = {black}, 
mark  = {o},
mark options = {solid},
style = {solid},
line width = 1pt] table {figs/datFiles/fig3-line2.dat};

\addlegendentry{Proposed Method - $ L_{\cellSubsetIdx} = 1 $}

\addplot [ color = {magenta}, 
mark  = {square},
mark options = {solid},
style = {solid},
line width = 1pt] table {figs/datFiles/fig3-line3.dat};

\addlegendentry{Proposed Method - Staggered Pilot}

\addplot [ color = {blue}, 
           mark  = {asterisk},
           mark options = {solid},
           style = {solid},
           line width = 1pt] table {figs/datFiles/fig3-line1.dat};

\end{axis}
\end{tikzpicture}

%% file: figs/fig4.tex
\begin{tikzpicture}[scale = 1]

\pgfplotsset{every axis legend/.append style={at={(0.35,0.25)},anchor=west},legend cell align = {left}}
        
\pgfplotsset{%
tick label style={font=\Large},
title style={font=\normalsize,align=center},
label style={font=\Large,align=center},
legend style={font=\Large}
            }

\begin{axis}[%
width=4.822in,
height=3.703in,
at={(0.809in,0.513in)},
xlabel = {$ \numCoherenceBlock $},
ylabel = {Avg. Sum Rate},
xmin = 0,
xtick = {0,500,1000,1500,2000,2500},
xmax = 2500,
ymin = 17,
ymax = 23,
cycle list name = {color},
grid = major,
scale only axis,
scale = 1]

\addlegendentry{LMMSE - Staggered Pilot}

\addplot [ color = {blue}, 
mark  = {none},
mark options = {solid},
style = {dashed},
line width = 1pt] table {figs/datFiles/fig4-line2.dat};

\addlegendentry{LMMSE - Regular Pilot}

\addplot [ color = {black}, 
mark  = {none},
mark options = {solid},
style = {solid},
line width = 1pt] table {figs/datFiles/fig4-line3.dat};

\addlegendentry{Proposed Method - Staggered Pilot}

\addplot [ color = {blue}, 
           mark  = {asterisk},
           mark options = {solid},
           style = {solid},
           line width = 1pt] table {figs/datFiles/fig4-line1.dat};

\addlegendentry{Method in \cite{bjornson2016imperfect}}

\addplot [ color = {black}, 
           mark  = {o},
           mark options = {solid},
           style = {solid},
           line width = 1pt] table {figs/datFiles/fig4-line4.dat};

\addlegendentry{LS}

\addplot [ color = {cyan}, 
mark  = {diamond},
mark options = {solid},
style = {solid},
line width = 1pt] table {figs/datFiles/fig4-line5.dat};

\end{axis}
\end{tikzpicture}

%% file: texFiles/pilotType.tex
\section{Estimating Covariance Matrices Using Staggered Pilots}
The proposed method in Section \ref{sec:proposedMethod} can be employed with either regular, superimposed, or staggered pilots. However, we will restrict our attention in this section to staggered pilots since it is a particular case of superimposed pilots and provides an additional degree of freedom by allowing the pilot and data powers to be varied \cite{mahyiddin2015performance,kong2016Channel,upadhya2017downlink}. The estimated channel will be used in a regularized zero-forcing (RZF) combiner for data detection, which is given as
\begin{align}
	\rxCombiner_{\cellIdx\userIdx}
	\!
	=
	\!
	\lrc{
		\!
			\frac{\sigma^2}{\ulTotPower{}{}} \eye 
			+
			\sum\limits_{\sIdxTwo=0}^{K-1} 
			\widehat{\mbf{h}}_{\cellIdx\cellIdx\sIdxTwo}^{\mathrm{LMMSE}}	
			\lrc{\widehat{\mbf{h}}_{\cellIdx\cellIdx\sIdxTwo}^{\mathrm{LMMSE}}}^H
			}^{\!\!-1}
			\!\!\!\!\!
	\widehat{\mbf{h}}_{\cellIdx\cellIdx\userIdx}^{\mathrm{LMMSE}}\;.
\end{align}

For covariance matrix estimation using staggered or regular pilots, each user transmits two $ K $ length pilot sequences within a coherence block, with the second pilot sequence multiplied by a random phase-shift. For each pilot sequence, $ \mbf{P} = \mathrm{blkdiag}\lrf{\mbf{\Phi},\ldots,\mbf{\Phi}} $ where $ \mbf{\Phi} \in \mathbb{C}^{K\times K} $. A lower bound on the capacity for user $ \lrc{0,\userIdx} $ can be obtained as
\begin{align}
& R \triangleq	\prelogFactor_1
\logCapacity
\lrc{
	\frac{s_{\userIdx}}{
		\bar{i}_{\userIdx}}
}
+
\prelogFactor_2
\logCapacity
\lrc{
	\frac{s_{\userIdx}}{\tilde{i}_{\userIdx}}
}
\label{eqn:staggPilotAchRate}
\end{align}
where $ \logCapacity\lrc{x} \triangleq\log_2\lrc{1+x} $, and
\begin{align}
&\prelogFactor_1
	\triangleq
	\frac{\lrc{L-1}K}{\ulDuration\numStationaryBlock}\lrc{\numStationaryBlock+\numCoherenceBlock}
;\!\!\!\quad
\prelogFactor_2
\triangleq
	\lrc{\!1\! -\! \frac{LK}{\ulDuration}\! - \frac{\numCoherenceBlock LK}{\ulDuration\numStationaryBlock}}
\\
&s_{\userIdx}
\triangleq
\rhoD{}{}^2
|		
\expectation
\lrf{
	\rxCombiner_{0\userIdx}^H
	\mbf{h}_{00\userIdx}
}|^2
\\	
&\bar{i}_{\userIdx}
\triangleq
\sum\limits_{\sIdxOne=0}^{L-1}
\sum\limits_{\sIdxTwo=0}^{K-1}	
\dataPilotPower_{\sIdxOne}
\expectation
\lrf{|\rxCombiner_{0\userIdx}^H \mbf{h}_{0\sIdxOne\sIdxTwo}|^2}	
-
\rhoD{}{}^2
|		
\expectation
\lrf{
	\rxCombiner_{0\userIdx}^H
	\mbf{h}_{00\userIdx}
}|^2			
\nonumber
\\
&	
+
\frac{\sigma^2}{\ulTotPower{}{}}
\expectation
\lrf{
	\|\rxCombiner_{0\userIdx}\|^2
}
+
\frac{\lrc{	\rhoP{}{}^2 - \rhoD{}{}^2}}
{
	\lrc{|\cellSubset_{0}|-1}}
\sum\limits_{\cellSubset_{0}\ni\cellIdx\neq 0}
\sum\limits_{\sIdxTwo=0}^{K-1}	
\expectation
\lrf{
	\left|
	\rxCombiner_{0\userIdx}^H
	\mbf{h}_{0\cellIdx\sIdxTwo}
	\right|^2
}
\\
&\tilde{i}_{\userIdx}
\triangleq 
\sum\limits_{\sIdxOne=0}^{L-1}
\sum\limits_{\sIdxTwo=0}^{K-1}	
\dataPilotPower_{\sIdxOne}
\expectation
\lrf{|\rxCombiner_{0\userIdx}^H \mbf{h}_{0\sIdxOne\sIdxTwo}|^2}	
+
\frac{\sigma^2}{\ulTotPower{}{}}
\expectation
\lrf{
	\|\rxCombiner_{0\userIdx}\|^2
}
\nonumber
\\
&-
\rhoD{}{}^2
|		
\expectation
\lrf{
	\rxCombiner_{0\userIdx}^H
	\mbf{h}_{00\userIdx}
}|^2\;.			
\end{align}
Here $ \dataPilotPower_{\sIdxOne} \triangleq \lrc{\rhoD{}{}^2\indicator{\sIdxOne\in\cellSubset_{0}} + \max\lrf{\rhoD{}{}^2,\rhoP{}{}^2}\indicator{\sIdxOne\notin\cellSubset_{0}}} $. The derivation of \eqref{eqn:staggPilotAchRate} is detailed in Appendix \ref{appdx:staggPilotAchRate}%
. Defining $ \errorCovarianceMatrix_{\vIdxOne\cellIdx\userIdx} \triangleq \covarianceMatrix_{\vIdxOne\cellIdx\userIdx} - \widehat{\covarianceMatrix}_{\vIdxOne\cellIdx\userIdx} $, the MSE expressions of the covariance matrix estimates can be obtained in a straightforward manner as
\begin{align}
\expectation\lrf{\|\errorCovarianceMatrix_{\vIdxOne\cellIdx\userIdx}\|^2_F}
=
\frac{1}{\numCoherenceBlock}
\sum\limits_{\sIdxThree=0}^{M-1}
\sum\limits_{\sIdxFour=0}^{M-1}
\lrc{
	g_{\sIdxThree}
	g_{\sIdxFour}
	+
	f_{\sIdxThree\sIdxFour}
}
\end{align}
where
\begin{align}
	&f_{\sIdxThree\sIdxFour}
	\!\triangleq
	\!\!
	\sum\limits_{\cellSubset_{\cellIdx}\ni\sIdxOne\neq\cellIdx}
	\!
	\sum\limits_{\sIdxTwo=0}^{K-1}
	\!\!
	\frac{\rhoD{}{}^4}{K^2\rhoP{}{}^4}
	|\lrs{\covarianceMatrix_{\sIdxOne\sIdxTwo}}_{\sIdxThree\sIdxFour}|^2
	+
	\left|
	\frac{1}{K\rhoP{}{}^2}
	\!
	\sum\limits_{\sIdxOne\in\cellSubsetComplement{\cellIdx}}
	\sum\limits_{\sIdxTwo=0}^{K-1}
	\lrs{\covarianceMatrix_{\sIdxOne\sIdxTwo}}_{\sIdxThree\sIdxFour}
	\right|^2
	\\
&g_{\sIdxThree}
\triangleq 
\lrs{\covarianceMatrix_{\cellIdx\userIdx}}_{\sIdxThree\sIdxThree}
+
\sum\limits_{\cellSubset_{\cellIdx}\ni\sIdxOne\neq\cellIdx}
\sum\limits_{\sIdxTwo=0}^{K-1}
\frac{\rhoD{}{}^2}{K\rhoP{}{}^2}
|\lrs{\covarianceMatrix_{\sIdxOne\sIdxTwo}}_{\sIdxThree\sIdxThree}|^2
\nonumber
\\
&+
\sum\limits_{\sIdxOne\in\cellSubsetComplement{\cellIdx}}
\sum\limits_{\sIdxTwo=0}^{K-1}
\frac{1}{K\rhoP{}{}^2}
|\lrs{\covarianceMatrix_{\sIdxOne\sIdxTwo}}_{\sIdxThree\sIdxThree}|^2
+
\frac{\sigma^2}{K\rhoP{}{}^2\ulTotPower{}{}}\;.
\end{align}

%% file: texFiles/simulationResults.tex
\section{Simulation Results}
\label{sec:simulationResults}
We compare the normalized MSE of the channel estimate and the achievable rate of the proposed method with that in \cite{bjornson2016imperfect}.\footnote{The code that reproduces the results in this section is available at \mbox{\url{https://github.com/karthikupadhya/covarianceEstimation-massiveMIMO}}.} Both the methods are simulated for one tier of $ L=7 $ hexagonal cells with the BSs separated by $ 300 $m. The performance of the methods are evaluated for the users in the central cell. The  SNR at the receiver is $ 78.6 - 37.6\log_{10}d $, where $ d $ is the distance from the BS in metres. The channel statistics are assumed to be constant over $ \numStationaryBlock = 25000 $ coherence blocks. The received paths from a user are assumed to be uniformly distributed over an angular spread of $ 20^{\circ} $, with mean angle of arrival given by the geographical locations of the users. 
In all the plots, the performance metrics are plotted against $ \numCoherenceBlock $. Then, $ \numCoherenceBlock = \nQ $ ensures that the same number of coherence blocks are utilized for estimating the covariance matrix for both the proposed method and \cite{bjornson2016imperfect}. For the method in \cite{bjornson2016imperfect}, $ \nR $ is chosen as $ \nR = \nQ / 10 $. The normalized MSE is defined as $ \expectation\lrf{\|\widehat{\mbf{h}}_{\vIdxOne\cellIdx\userIdx}^{\mathrm{LMMSE}} - \mbf{h}_{\vIdxOne\cellIdx\userIdx}\|^2 / \mathrm{trace}\lrf{\covarianceMatrix_{\vIdxOne\cellIdx\userIdx}}} $. For the proposed method, $ \rhoD{}{}^2 = \rhoP{}{}^2 = 1 $. 

In Figs. \ref{fig:mseVsN}, \ref{fig:rateVsN}, and \ref{fig:covMatMseVsN}, the performance of the proposed method is compared with that in \cite{bjornson2016imperfect}. In order to compare only the covariance matrix estimates, the proposed method utilizes staggered pilots for $ \numCoherenceBlock $ coherence blocks and regular pilots for the remaining $ \numStationaryBlock-\numCoherenceBlock $ blocks. Consequently, $ \widehat{\sumCovarianceMatrix}_{\vIdxOne\userIdx} $ is estimated for the proposed method as $ \widehat{\sumCovarianceMatrix}_{\vIdxOne\userIdx} = \sum\limits_{\sIdxOne=0}^{L-1} \widehat{\covarianceMatrix}_{\vIdxOne\sIdxOne\userIdx} + (\sigma^2/K\ulTotPower{}{}) \eye $. Note that we have used the same simulation setup as in \cite{bjornson2016imperfect} in which $ \ulDuration = 100 $ symbols, each BS has $ M=100 $ antennas and contains $ K=10 $ users in its cell which are equispaced on a circle of radius $ 120 $m from the BS. In addition, for staggered pilots $ L_t $ is chosen as $ 7 $. For the sake of simplicity, $ L_t = 1 $ is simulated using regular pilots, although the proposed method would still work if the pilot and data transmissions of different cells would overlap. From Fig. \ref{fig:covMatMseVsN}, it can be seen the MSE of the covariance matrix is significantly lower for the proposed method. Consequently, in Figs. \ref{fig:mseVsN} and \ref{fig:rateVsN}, the MSE and sum rate performance of the proposed method is significantly better than the method in \cite{bjornson2016imperfect}, despite not requiring all users to transmit the pilots over the same set of symbols.

In Fig.  \ref{fig:rateVsNUniform}, $ K=5 $ users are uniformly distributed across the entire cell. Each BS has $ M = 50 $ antennas and the UL time slot has $ \ulDuration = 100 $ symbols. With the LMMSE method, the sum-rate with regular pilots is marginally higher than that for staggered pilots. However, the proposed method offers a higher throughput in comparison with the method in \cite{bjornson2016imperfect}. Note that the pre-log factor contributes to the small difference between the achievable rates of the LMMSE and estimated covariance matrices.

%% file: texFiles/conclusion.tex
\section{Conclusion}
\label{sec:conclusion}
We proposed a novel pilot structure for estimating the covariance matrix in the presence of pilot contamination, which has the advantage of not requiring simultaneous UL pilot transmissions. Using the proposed method along with staggered pilots, we showed that the proposed method offers a higher UL throughput and lower MSE than existing schemes. 

The performance of the proposed method could be further improved by optimizing $ \rhoD{}{} $ and $ \rhoP{}{} $, which we leave as a problem for future research.

%% file: texFiles/appendix.tex
\appendix
\subsection{Lower Bound on the Channel Capacity for Staggered Pilots}
\label{appdx:staggPilotAchRate}
We consider an arbitrary cell with index $ 0 $ and with the associated subset of cells $ \cellSubset_{0} $. The UL symbols are indexed from $ 1 $  to $ \ulDuration $ and without loss of generality, we assume that the symbols with index $ \cellIdx K + 1 $ to $ (\cellIdx+1) K $ are used for pilot transmission by the users in cell $ \cellIdx \in \cellSubset_{0} $. We also assume that the users in $ \cellSubsetComplement{0} $ transmit their symbols with power $ \ulTotPower{}{}\max\lrf{\rhoD{}{}^2,\lambda{}{}^2} $. This is a worst case scenario since it causes the maximum interference to the users in cell $ 0 $. If $ 1\leq\dataIdx\leq K $, the received observation vector at BS $ 0 $ when users in cell $ \cellIdx \in \cellSubset_{0} $ are transmitting pilots can be written as
\begin{align}
	&\lrs{\mbf{y}^{\lrc{\cellIdx}}}_{\dataIdx}
	\!
	=
	\!
	\sum\limits_{\sIdxTwo=0}^{K-1}
	\sqrt{\ulTotPower{}{}}
	\rhoD{}{}
	\mbf{h}_{00\sIdxTwo}
	\!
	\lrs{\mbf{x}_{0\sIdxTwo}^{(\cellIdx)}}_{\dataIdx}
	\!
	+
	\!\!
	\sum\limits_{\cellSubset_{0} \ni \sIdxOne \neq 0 }
	\sum\limits_{\sIdxTwo=0}^{K-1}
	\sqrt{\ulTotPower{}{}}
	\rhoD{}{}	
	\mbf{h}_{0\sIdxOne\sIdxTwo}
	\lrs{\mbf{x}_{\sIdxOne\sIdxTwo}^{(\cellIdx)}}_{\dataIdx}
	\nonumber\\
	&
	+
	\!
	\sum\limits_{\sIdxTwo=0}^{K-1}
	\!
	\sqrt{\ulTotPower{}{}}
	\rhoP{}{}	
	\mbf{h}_{0\cellIdx\sIdxTwo}
	\!
	\lrs{\pmb{\phi}_{\cellIdx\sIdxTwo}^T}_{\dataIdx}
	\!
	+
	\!
	\sum\limits_{\sIdxOne \notin \cellSubset_{0}}
	\sum\limits_{\sIdxTwo=0}^{K-1}
	\!
	\sqrt{\ulTotPower{}{}}
	\max\lrf{\rhoD{}{},\rhoP{}{}}
	\mbf{h}_{0\sIdxOne\sIdxTwo}
	\lrs{\mbf{x}_{\sIdxOne\sIdxTwo}^{(\cellIdx)}}_{\dataIdx}
	\nonumber
	\\
	&+
	\lrs{\mbf{w}^{\lrc{\cellIdx}}}_{\dataIdx}
	\label{eqn:staggPilotRxObsData+Pilot}
\end{align}
The output of the combiner $ \rxCombiner_{0\userIdx} $ can be written as 
\begin{align}
	&\rxCombiner_{0\userIdx}^H
	\lrs{\mbf{y}^{\lrc{\cellIdx}}}_{\dataIdx}
	=
	\sqrt{\ulTotPower{}{}}
	\rhoD{}{}
	\expectation
	\lrf{
			\rxCombiner_{0\userIdx}^H
			\mbf{h}_{00\userIdx}
		}
	\lrs{\mbf{x}_{0\userIdx}^{(\cellIdx)}}_{\dataIdx}
	\nonumber
	\\
	&
	+
	\sqrt{\ulTotPower{}{}}
	\rhoD{}{}
	\lrc{
			\rxCombiner_{0\userIdx}^H
			\mbf{h}_{00\userIdx}
			-
			\expectation
			\lrf{
				\rxCombiner_{0\userIdx}^H
				\mbf{h}_{00\userIdx}
			}		
		}
	\lrs{\mbf{x}_{0\userIdx}^{(\cellIdx)}}_{\dataIdx}
	+	
	e^{\lrc{\cellIdx}}_{\dataIdx}
	\label{eqn:staggPilotCombinerOuputData+Pilot}
\end{align}
where
\begin{align}
	&e^{\lrc{\cellIdx}}_{\dataIdx}
	\triangleq
	\sum\limits_{\sIdxTwo\neq\userIdx}
	\sqrt{\ulTotPower{}{}}
	\rhoD{}{}
	\rxCombiner_{0\userIdx}^H
	\mbf{h}_{00\sIdxTwo}
	\lrs{\mbf{x}_{0\sIdxTwo}^{(\cellIdx)}}_{\dataIdx}	
	+
	\sum\limits_{\sIdxTwo=0}^{K-1}
	\sqrt{\ulTotPower{}{}}
	\rhoP{}{}
	\rxCombiner_{0\userIdx}^H
	\mbf{h}_{0\cellIdx\sIdxTwo}
	\lrs{\pmb{\phi}_{\cellIdx\sIdxTwo}^T}_{\dataIdx}	
	\nonumber\\
	&
	+
	\sum\limits_{\sIdxOne \notin \cellSubset_{0}}
	\sum\limits_{\sIdxTwo=0}^{K-1}
	\sqrt{\ulTotPower{}{}}
	\max\lrf{\rhoD{}{},\rhoP{}{}}
	\rxCombiner_{0\userIdx}^H
	\mbf{h}_{0\sIdxOne\sIdxTwo}
	\lrs{\mbf{x}_{\sIdxOne\sIdxTwo}^{(\cellIdx)}}_{\dataIdx}
	\nonumber
	\\
	&
	+
	\sum\limits_{\cellSubset_{0} \ni \sIdxOne \neq 0 }
	\sum\limits_{\sIdxTwo=0}^{K-1}
	\sqrt{\ulTotPower{}{}}
	\rhoD{}{}	
	\rxCombiner_{0\userIdx}^H
	\mbf{h}_{0\sIdxOne\sIdxTwo}
	\lrs{\mbf{x}_{\sIdxOne\sIdxTwo}^{(\cellIdx)}}_{\dataIdx}	
	+
	\rxCombiner_{0\userIdx}^H
	\mbf{w}\;.
\end{align}
Noting in \eqref{eqn:staggPilotCombinerOuputData+Pilot} that the first term is uncorrelated with the subsequent terms, the signal and interference powers can be obtained as
\begin{align}
	&s_{\userIdx}
	= \ulTotPower{}{}
	\rhoD{}{}^2
	|		
	\expectation
	\lrf{
		\rxCombiner_{0\userIdx}^H
		\mbf{h}_{00\userIdx}
	}|^2
	\\	
	&\lrs{i^{(\cellIdx)}_{\userIdx}}_{\dataIdx} 
	\!
	= 
	\ulTotPower{}{}		
	\sum\limits_{\sIdxOne\neq\cellIdx}
	\sum\limits_{\sIdxTwo=0}^{K-1}	
	\dataPilotPower_{\sIdxOne}	
	\expectation	
	\lrf{|\rxCombiner_{0\userIdx}^H \mbf{h}_{0\sIdxOne\sIdxTwo}|^2	}	
	\!
	-
	\ulTotPower{}{}
	\rhoD{}{}^2
	|		
	\expectation
	\lrf{
		\rxCombiner_{0\userIdx}^H
		\mbf{h}_{00\userIdx}
		\!
	}\!|^2			
	\nonumber
	\\
	&	
	+
	\sigma^2
	\expectation
	\lrf{
	\|\rxCombiner_{0\userIdx}\|^2
	}
	+
	\ulTotPower{}{}
	\rhoP{}{}^2
	\expectation
	\lrf{
	\left|
	\sum\limits_{\sIdxTwo=0}^{K-1}	
	\rxCombiner_{0\userIdx}^H
	\mbf{h}_{0\cellIdx\sIdxTwo}
	\lrs{\pmb{\phi}_{\cellIdx\sIdxTwo}^T}_{\dataIdx}
	\right|^2
	}
	\label{eqn:interfPowerOnlyPilot}
	\end{align}
Now, if $ \ulDuration > L_{0} K $, there are $ \ulDuration - L_{0} K $ symbols in the UL time-slot wherein none of the users in the $ L_{0} $ cells are transmitting UL pilots. The interference power in these symbols can be obtained in a straightforward manner as
\begin{align}
	&\tilde{i}_{\userIdx}
	= 
	\ulTotPower{}{}		
	\sum\limits_{\sIdxOne=0}^{L-1}
	\sum\limits_{\sIdxTwo=0}^{K-1}	
	\dataPilotPower_{\sIdxOne}	
	\expectation
	\lrf{|\rxCombiner_{0\userIdx}^H \mbf{h}_{0\sIdxOne\sIdxTwo}|^2	}	
	-
	\ulTotPower{}{}
	\rhoD{}{}^2
	|		
	\expectation
	\lrf{
		\rxCombiner_{0\userIdx}^H
		\mbf{h}_{00\userIdx}
	}|^2				
	\nonumber
	\\
	&	
	+
	\sigma^2
	\expectation
	\lrf{
		\|\rxCombiner_{0\userIdx}\|^2
	}
	\label{eqn:interfPowerOnlyData} \;.
\end{align}
Therefore, a lower bound on the achievable rate can be obtained as
\begin{align}
	&\achRate_{0\userIdx}
	=
	\frac{\numCoherenceBlock}{\ulDuration\numStationaryBlock}
	\left[
	\sum\limits_{\cellSubset_{0}\ni\cellIdx\neq 0}
	\sum\limits_{\dataIdx=1}^{2K}
	\logCapacity
	\lrc{
			\frac{s_{\userIdx}}{\lrs{i^{(\cellIdx)}_{\userIdx}}_{\dataIdx} }
		}
	+
	\numExtraSymbol_{2K}
	\logCapacity
	\lrc{
		\frac{s_{\userIdx}}{\tilde{i}_{\userIdx}}
	}
	\right]
	\nonumber
	\\
	&+
	\frac{\lrc{\numStationaryBlock-\numCoherenceBlock}}{\ulDuration\numStationaryBlock}
	\left[
	\sum\limits_{\cellSubset_{0}\ni\cellIdx\neq 0}
	\sum\limits_{\dataIdx=1}^{K}
	\logCapacity
	\lrc{
		\frac{s_{\userIdx}}{\lrs{i^{(\cellIdx)}_{\userIdx}}_{\dataIdx} }
	}	
	+
	\numExtraSymbol_{K}
	\logCapacity
	\lrc{
		\frac{s_{\userIdx}}{\tilde{i}_{\userIdx}}
	}
	\right]
	\label{eqn:staggAchRate}
\end{align}
where $ \numExtraSymbol_{x}= \max\lrf{\ulDuration-L_{0}x,0} $.
The first two terms correspond to the throughput in the $ \numCoherenceBlock $ coherence blocks in which $ 2K $ pilots are transmitted by each user for estimating the covariance matrix and the last two terms correspond to the remaining $ \numStationaryBlock-\numCoherenceBlock $ blocks where only $ K $ pilots are used for estimating the channel. Since $ \log_2\lrc{1 + 1/x} $ is convex in $ x $, using Jensen's inequality a lower bound on \eqref{eqn:staggAchRate} can be obtained as
\begin{align}
	&\achRate_{0\userIdx}
	=
	\prelogFactor_1
	\logCapacity
	\lrc{
		\frac{s_{\userIdx}}{
			\bar{i}_{\userIdx}}
	}
	+
	\prelogFactor_2
	\logCapacity
	\lrc{
		\frac{s_{\userIdx}}{\tilde{i}_{\userIdx}}
	}
\end{align}
where
\begin{align}
	\bar{i}_{\userIdx}
	&=
	\frac{1}{2\lrc{|\cellSubset_{0}|-1}}
\sum\limits_{\cellSubset_{0}\ni\cellIdx\neq 0}
\sum\limits_{\dataIdx=1}^{K}\lrs{i^{(\cellIdx)}_{\userIdx}}_{\dataIdx}
	\\	
	&=
	\ulTotPower{}{}		
	\sum\limits_{\sIdxOne=0}^{L-1}
	\sum\limits_{\sIdxTwo=0}^{K-1}	
	\dataPilotPower_{\sIdxOne}	
	\expectation
	\lrf{|\rxCombiner_{0\userIdx}^H \mbf{h}_{0\sIdxOne\sIdxTwo}|^2}	
	-
	\ulTotPower{}{}
	\rhoD{}{}^2
	|		
	\expectation
	\lrf{
		\rxCombiner_{0\userIdx}^H
		\mbf{h}_{00\userIdx}
	}|^2			
	\nonumber
	\\
	&
	\!\!\!\!	
	+
	\sigma^2
	\expectation
	\lrf{
		\|\rxCombiner_{0\userIdx}\|^2
	}
	+
	\ulTotPower{}{}	
	\frac{\lrc{\rhoP{}{}^2 - \rhoD{}{}^2}}
	{
		\lrc{|\cellSubset_{0}|-1}}
	\sum\limits_{\cellSubset_{0}\ni\cellIdx\neq 0}
		\sum\limits_{\sIdxTwo=0}^{K-1}	
	\expectation
	\lrf{
		\left|
		\rxCombiner_{0\userIdx}^H
		\mbf{h}_{0\cellIdx\sIdxTwo}
		\right|^2
	}
	\\
	\prelogFactor_1
	&=
	\frac{\lrc{L-1}K}{\ulDuration\numStationaryBlock}\lrc{\numStationaryBlock+\numCoherenceBlock}
	\\
	\prelogFactor_2
	&=
	\lrc{1 - \frac{LK}{\ulDuration} - \frac{\numCoherenceBlock LK}{\ulDuration\numStationaryBlock}}.
\end{align}

%% file: Main.bbl
\begin{thebibliography}{10}
\providecommand{\url}[1]{#1}
\csname url@samestyle\endcsname
\providecommand{\newblock}{\relax}
\providecommand{\bibinfo}[2]{#2}
\providecommand{\BIBentrySTDinterwordspacing}{\spaceskip=0pt\relax}
\providecommand{\BIBentryALTinterwordstretchfactor}{4}
\providecommand{\BIBentryALTinterwordspacing}{\spaceskip=\fontdimen2\font plus
\BIBentryALTinterwordstretchfactor\fontdimen3\font minus
  \fontdimen4\font\relax}
\providecommand{\BIBforeignlanguage}[2]{{%
\expandafter\ifx\csname l@#1\endcsname\relax
\typeout{** WARNING: IEEEtran.bst: No hyphenation pattern has been}%
\typeout{** loaded for the language `#1'. Using the pattern for}%
\typeout{** the default language instead.}%
\else
\language=\csname l@#1\endcsname
\fi
#2}}
\providecommand{\BIBdecl}{\relax}
\BIBdecl

\bibitem{Marzetta2010Noncooperative}
T.~Marzetta, ``Noncooperative cellular wireless with unlimited numbers of base
  station antennas,'' \emph{{IEEE} Trans. Wireless Commun.}, vol.~9, no.~11,
  pp. 3590--3600, Nov. 2010.

\bibitem{larsson2014massive}
E.~Larsson, O.~Edfors, F.~Tufvesson, and T.~Marzetta, ``Massive {MIMO} for next
  generation wireless systems,'' \emph{{IEEE} Commun. Mag.}, vol.~52, no.~2,
  pp. 186--195, Feb. 2014.

\bibitem{lulu2014anoverview}
L.~Lu, G.~Li, A.~Swindlehurst, A.~Ashikhmin, and R.~Zhang, ``An overview of
  massive {MIMO}: Benefits and challenges,'' \emph{{IEEE} J. Sel. Topics Signal
  Process.}, vol.~8, no.~5, pp. 742--758, Oct. 2014.

\bibitem{rusek2013scaling}
F.~Rusek, D.~Persson, B.~K. Lau, E.~Larsson, T.~Marzetta, O.~Edfors, and
  F.~Tufvesson, ``Scaling up {MIMO}: Opportunities and challenges with very
  large arrays,'' \emph{{IEEE} Signal Process. Mag.}, vol.~30, no.~1, pp.
  40--60, Jan. 2013.

\bibitem{yang2013performance}
H.~Yang and T.~Marzetta, ``Performance of conjugate and zero-forcing
  beamforming in large-scale antenna systems,'' \emph{{IEEE} J. Sel. Areas
  Commun.}, vol.~31, no.~2, pp. 172--179, Feb. 2013.

\bibitem{hoydis2013massive}
J.~Hoydis, S.~ten Brink, and M.~Debbah, ``Massive {MIMO} in the {UL/DL} of
  cellular networks: How many antennas do we need?'' \emph{{IEEE} J. Sel. Areas
  Commun.}, vol.~31, no.~2, pp. 160--171, Feb. 2013.

\bibitem{Muller2014Blind}
R.~Muller, L.~Cottatellucci, and M.~Vehkapera, ``Blind pilot decontamination,''
  \emph{{IEEE} J. Sel. Topics Signal Process.}, vol.~8, no.~5, pp. 773--786,
  Oct. 2014.

\bibitem{bjornson2015massive}
E.~Bj\"ornson, E.~G. Larsson, and M.~Debbah, ``Massive {MIMO} for maximal
  spectral efficiency: How many users and pilots should be allocated?''
  \emph{{IEEE} Trans. Wireless Commun.}, vol.~15, no.~2, pp. 1293--1308, Feb.
  2016.

\bibitem{Yin2013Coordinated}
H.~Yin, D.~Gesbert, M.~Filippou, and Y.~Liu, ``A coordinated approach to
  channel estimation in large-scale multiple-antenna systems,'' \emph{{IEEE} J.
  Sel. Areas Commun.}, vol.~31, no.~2, pp. 264--273, Feb. 2013.

\bibitem{upadhya2015superimposed}
K.~Upadhya, S.~A. Vorobyov, and M.~Vehkapera, ``Superimposed pilots: An
  alternative pilot structure to mitigate pilot contamination in massive
  {MIMO},'' in \emph{Proc. IEEE Int. Conf. on Acoustics, Speech and Signal
  Processing (ICASSP)}, Shanghai, Mar. 2016, pp. 3366--3370.

\bibitem{upadhya2016superimposed}
------, ``Superimposed pilots are superior for mitigating pilot contamination
  in massive {MIMO},'' \emph{{IEEE} Trans. Signal Process.}, vol.~65, no.~11,
  pp. 2917--2932, Jun. 2017.

\bibitem{jose2011pilot}
J.~Jose, A.~Ashikhmin, T.~Marzetta, and S.~Vishwanath, ``Pilot contamination
  and precoding in multi-cell {TDD} systems,'' \emph{{IEEE} Trans. Wireless
  Commun.}, vol.~10, no.~8, pp. 2640--2651, Aug. 2011.

\bibitem{bjornson2017unlimited}
\BIBentryALTinterwordspacing
E.~Bj\"{o}rnson, J.~Hoydis, and L.~Sanguinetti, ``Massive {MIMO} has unlimited
  capacity,'' \emph{IEEE Trans. Wireless Commun., To be published}. [Online].
  Available: \url{http://arxiv.org/abs/1705.00538}
\BIBentrySTDinterwordspacing

\bibitem{bjornson2016imperfect}
E.~Bj\"{o}rnson, L.~Sanguinetti, and M.~Debbah, ``Massive {MIMO} with imperfect
  channel covariance information,'' in \emph{Proc. Asilomar Conf. on Signals,
  Systems and Computers}, Nov. 2016, pp. 974--978.

\bibitem{neumann2017jointCovariance}
D.~Neumann, K.~Shibli, M.~Joham, and W.~Utschick, ``Joint covariance matrix
  estimation and pilot allocation in massive {MIMO} systems,'' in \emph{Proc.
  IEEE Int. Conf. on Communications (ICC)}, May 2017, pp. 1--6.

\bibitem{caire2017massive}
\BIBentryALTinterwordspacing
S.~Haghighatshoar and G.~Caire, ``Massive {MIMO} pilot decontamination and
  channel interpolation via wideband sparse channel estimation,'' \emph{arXiv},
  2017. [Online]. Available: \url{http://arxiv.org/abs/1702.07207}
\BIBentrySTDinterwordspacing

\bibitem{bjornson2017pilot}
E.~Bj\"{o}rnson, J.~Hoydis, and L.~Sanguinetti, ``Pilot contamination is not a
  fundamental asymptotic limitation in massive {MIMO},'' in \emph{Proc. IEEE
  Int. Conf. on Communications (ICC)}, May 2017, pp. 1--6.

\bibitem{mahyiddin2015performance}
W.~A. W.~M. Mahyiddin, P.~A. Martin, and P.~J. Smith, ``Performance of
  synchronized and unsynchronized pilots in finite massive {MIMO} systems,''
  \emph{{IEEE} Trans. Wireless Commun.}, vol.~14, no.~12, pp. 6763--6776, Dec.
  2015.

\bibitem{kong2016Channel}
D.~Kong, D.~Qu, K.~Luo, and T.~Jiang, ``Channel estimation under staggered
  frame structure for massive {MIMO} system,'' \emph{{IEEE} Trans. Wireless
  Commun.}, vol.~15, no.~2, pp. 1469--1479, Feb. 2016.

\bibitem{upadhya2017downlink}
\BIBentryALTinterwordspacing
K.~Upadhya, S.~A. Vorobyov, and M.~Vehkapera, ``Downlink performance of
  superimposed pilots in massive {MIMO} systems,'' \emph{Submitted to IEEE
  Trans. Wireless Commun.} [Online]. Available:
  \url{https://arxiv.org/pdf/1606.04476.pdf}
\BIBentrySTDinterwordspacing

\end{thebibliography}
